\documentclass{INTERSPEECH2023}

\interspeechcameraready

\usepackage[ruled,linesnumbered]{algorithm2e}
\usepackage{multirow}
\usepackage{hyperref}
\usepackage{cleveref}

\hypersetup{
    hidelinks,
    colorlinks=true,
    pdfborder={0 0 1},
    linkbordercolor={1 0 0},
}

\title{Enhance Temporal Relations in Audio Captioning with Sound Event Detection}
\name{Zeyu Xie, Xuenan Xu, Mengyue Wu$\dag$, Kai Yu$\dag$\thanks{$\dag$Mengyue Wu and Kai Yu are the corresponding authors.}}
\address{
  MoE Key Lab of Artificial Intelligence
   X-LANCE Lab, \\
   Department of Computer Science and Engineering
   AI Institute, \\
   Shanghai Jiao Tong University, Shanghai, China
}
\email{\{zeyu\_xie, wsntxxn, mengyuewu, kai.yu\}@sjtu.edu.cn}

\begin{document}

\maketitle
 
\begin{abstract}
Automated audio captioning aims at generating natural language descriptions for given audio clips, not only detecting and classifying sounds, but also summarizing the relationships between audio events. 
Recent research advances in audio captioning have introduced additional guidance to improve the accuracy of audio events in generated sentences.
However, temporal relations between audio events have received little attention while revealing complex relations is a key component in summarizing audio content.
Therefore, this paper aims to better capture temporal relationships in caption generation with sound event detection (SED), a task that locates events' timestamps.
We investigate the best approach to integrate temporal information in a captioning model and propose a temporal tag system to transform the timestamps into comprehensible relations.  
Results evaluated by the proposed temporal metrics suggest that great improvement is achieved in terms of temporal relation generation\footnote{The pre-trained model is available \href{https://github.com/wsntxxn/AudioCaption?tab=readme-ov-file\#temporal-sensitive-and-controllable-model}{here}.}.


\end{abstract}

\section{Introduction}

Increasing amount of research has shed light on machine perception of audio events, for instance label-wise classification and detection.
Recently automated audio captioning (AAC)~\cite{drossos2017automated} has gathered much attention due to its resemblance to human perception, which involves not only detecting and classifying sounds, but also summarizing the relationship between different audio events~\cite{wu2019audio}.
Over the last few years, AAC has witnessed remarkable advances in recent works.
The utilization of pre-trained audio classification and language generation models improve the captioning performance significantly~\cite{xu2021investigating,mei2021audio}.
The incorporation of semantic guidance (e.g., keywords~\cite{koizumi2020ntt,ye2021improving,eren2020semantic}, sound tags~\cite{gontier2021automated} or similar captions~\cite{koizumi2020audio}) and new loss functions~\cite{cakir2020multi,xu2021audio,liu2021cl4ac} are also hot topics.
While previous work endeavors to better detect audio events and improve caption quality, little attention is paid to summarizing relations between different sound events in a caption.
The current captioning model rarely outputs sentences involving temporal conjunctions like ``before'', ``after'' and ``followed by'' that suggest the sequential relations between events. 
A statistical examination on a well-performing AAC model~\cite{xu2021investigating} indicates that only 11.1\% generated captions include precise temporal relations. 

\begin{figure}[ht]
  \centering
  \includegraphics[width=\linewidth]{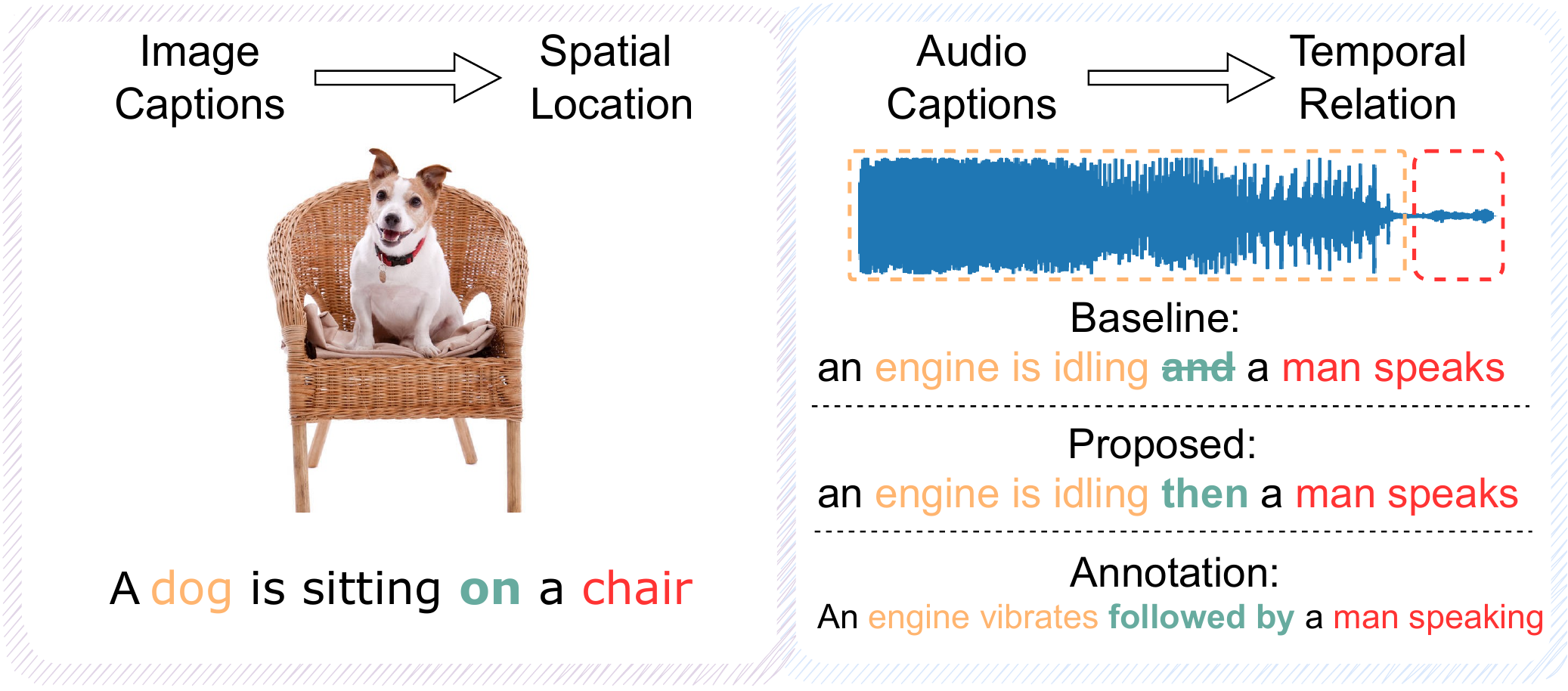}
  \caption{Expressions of relationships in image versus audio. Image pays more attention to spatial relations while audio focus on temporal relations.}
  \label{fig:image_vs_audio}
\end{figure}

Different from vision-based captioning where a plethora of spatial attributes can be extracted, audio events' relations are mainly focused on their time specificity as shown in \Cref{fig:image_vs_audio}.
Whether two audio events occur sequentially or simultaneously is important to understand the audio content correctly, which is as critical as whether two objects in an image are adjacent, stacked, or overlaid.

\begin{figure*}[ht]
  \centering
  \includegraphics[width=0.93\linewidth]{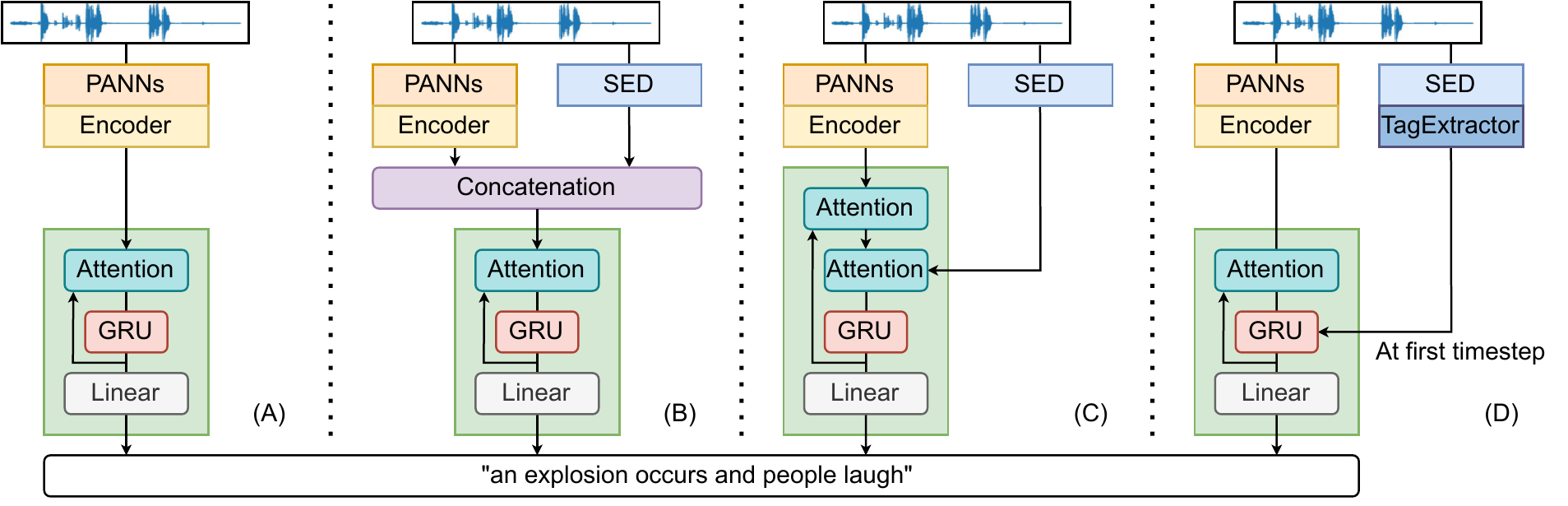}
  \caption{An overview of different AAC models.
(A) Baseline AAC model: the decoder generates captions solely based on audio embeddings;
(B) Cat-prob-AAC: audio embeddings and SED outputs are concatenated and used as the input to the decoder;
(C) Attn-prob-AAC: an attention mechanism is used to integrate SED outputs and decoder hidden states;
(D) Temp-tag-AAC: mimicking human judgment, tags are extracted and used as the input at the first timestep instead of $<$BOS$>$.}
  \label{fig:model}
\end{figure*}

Sound event detection (SED), a task to detect on- and off-sets of each sound event, on the other hand, provides extensive information on the temporal location of each event.
Previous works integrated SED outputs by direct concatenation to improve the overall quality and accuracy of generated captions~\cite{gontier2021automated,eren2021audio}.
However, whether such straightforward fusion methods can help a captioning model learn about temporal relations between events remains unexplored.
SED output contains information about the occurring probability of hundreds of sound events in each frame.
These redundant low-level features are difficult to align with the temporal conjunction words in a caption, making it difficult for the captioning model to leverage SED outputs.
In this work, we first directly integrate SED outputs by concatenation (\texttt{cat-prob-AAC}) and attention (\texttt{attn-prob-AAC}), to investigate the performance of direct SED integration methods.
The results demonstrate that such approaches bring little improvement in temporal relationship description accuracy.

Therefore, it is necessary to distill high-level, comprehensible temporal information from SED outputs, for a better alignment with audio caption content to mimic humans' temporal information processing procedure.
Inspired by this, we first analyse the current AAC data and propose a 4-scale temporal relation tagging system (i.e. simultaneous, sequential) based on human annotations.
A clear matching mechanism is further proposed to infer the temporal relations from SED outputs and align with the temporal tags.
Based on this, we propose a temporal tag-guided captioning system (\texttt{temp-tag-AAC}), which takes temporal tag guidance inferred from SED output that represents the complexity of temporal information to facilitate the model to generate captions with accurate temporal expressions.


To measure the quality of generated captions in terms of temporal relationship descriptions, we propose $\text{ACC}_{temp}$ and $\text{F1}_{temp}$.
Evaluated by these temporal-focused metrics and commonly-adopted captioning metrics (e.g., BLEU) indicate that temp-tag-AAC significantly outperforms the baseline model and the direct SED integration approach, especially in temporal relationship description accuracy.
Our contributions are summarized as follows:


\begin{enumerate}

\item Innovative utilization of SED to enhance the temporal information in AAC, with a temporal tag to better imitate humans' inference on temporal relations.

\item Metrics that are specifically designed to measure a system's capability in describing sound events' temporal relations.

\item Validation shows that the proposed temp-tag-AAC leverages SED outputs to significantly improves the accuracy of temporal expression as well as the caption quality. 

\end{enumerate}


\section{Temporal-Enhanced Captioning System}
This section illustrates our temporal-enhanced captioning system shown in \Cref{fig:model}, includes: 1) the baseline model for audio captioning; 2) the SED model that predicts the probability of events; 3) two direct approaches for integrating probability as temporal information; 4) proposed temp-tag-AAC approach. 

\begin{table*}[htpb]
\caption{Temporal Tags Extracted from Text (Captions) and Audio (SED Results), c.w. = Conjunction Words.} 
\label{tab_tamporal_tag}
\centering
    \begin{tabular}{c|c|c|c|c} 
    \toprule
    Temporal Tag & 0 & 1 & 2 & 3 \\
    \midrule
    Caption & No c.w. & Simultaneous c.w. & Sequential c.w. & More Complex Relations \\
    \hline
    SED & Only 1 Event & Simultaneous Events & Sequential Events & More Complex Events\\
    \bottomrule
    \end{tabular}
\end{table*}

\subsection{Baseline Approach}

The baseline framework follows an encoder-decoder architecture which achieves competitive performance in DCASE challenges~\cite{xu2022sjtu}.


\paragraph*{Audio Encoder} PANNs~\cite{kong2020panns} CNN14, a pre-trained convolutional neural network, is adopted to extract the feature from the input audio $\mathcal{A}$. 
We use a bidirectional gated recurrent unit (GRU) network as the audio encoder to transform the feature into an embedding sequence $\mathbf{e}^A \in \mathbb{R}^{T \times D}$.
The combination takes advantage of the pre-trained large model while setting some parameters trainable for adaptation to the target captioning task. 

\begin{equation}
  \mathbf{e}^A = \mathrm{\mathbf{Encoder}}(\mathrm{\mathbf{PANNs}}(\mathcal{A}))
  \label{eq_encoder}
\end{equation}

\paragraph*{Text Decoder} 
We use a unidirectional GRU as the text decoder to predict the caption word by word. 
At each timestep $n$, a context vector $\mathbf{c}$ is calculated by attention mechanism~\cite{bahdanau2015neural}, given $\mathbf{e}^A$ and the previous hidden state $\mathbf{h}_{n-1}$:

\begin{align}
    \begin{split}
    \alpha_{n,t} = \frac{\mathrm{exp(score}(\mathbf{h}_{n-1},\mathbf{e}_t^A))}{\sum_{t=1}^{T}{\mathrm{exp(score}(\mathbf{h}_{n-1},\mathbf{e}_t^A))}}\\
    \mathbf{c} = \mathrm{\mathbf{ATTN}}(\mathbf{h}_{n-1},\mathbf{e}^A) = \sum_{t=1}^{T}{\alpha_{n,t}\mathbf{e}_{t}}
    \label{eq:baseline_attn}
    \end{split}
\end{align}

Then the text decoder predicts the next word based on previously generated words $w_{0:n}$ and $\mathbf{c}$.
At the first timestep, $w_0$ is a special ``$<$BOS$>$'' token denoting the beginning of a sentence.





\subsection{SED Architecture}

To ensure the reliability of the SED results, we use a separately-trained SED model.
It adopts a convolutional recurrent neural network architecture with 8 convolutional layers attached by a BiGRU.
The convolution layers take a structure similar to the CNN10 in PANNs, with the difference that we use a downsampling ratio of 4 on the temporal axis.
Compared with other SED models provided in PANNs which typically utilize a downsampling ratio of 32, we keep a relatively high temporal resolution for more accurate SED.

Given an audio clip, the SED model outputs the predicted probability $\tilde{\mathbf{e}}^{S} \in \mathbb{R}^{\tilde{T} \times M}$, where $\tilde{T}$ and $M$ denote the sequence length and the number of sound event categories respectively.
Due to the higher resolution of the SED model, $\tilde{T} > T$.
The probability is temporally aligned to the audio embedding to obtain $\mathbf{e}^S \in \mathbb{R}^{T \times M}$ by pooling on every $\frac{\tilde{T}}{T}$ segments along the temporal axis.

\subsection{Direct SED Integration}

\paragraph*{Cat-prob-AAC} The probability is concatenated onto audio embedding, resulting in $\mathbf{e}^{A_{new}} \in \mathbb{R}^{T \times (D+M)} $, which is used as the input to the decoder instead of the original $\mathbf{e}^A$.

\begin{equation}
  \mathbf{e}^{A_{new}} = \mathrm{\mathbf{CONCAT}}(\mathbf{e}^A, \mathbf{e}^{S})
  \label{eq_concat}
\end{equation}

\paragraph*{Attn-prob-AAC}
\label{sec_atten} 
Another attention is used to integrate the probability and context vector $\mathbf{c}$ obtained from \Cref{eq:baseline_attn}.
The result is used as the input to the GRU instead of $\mathbf{c}$.

\begin{equation}
  \mathbf{c}^{new} = \mathrm{\mathbf{ATTN}}(\mathbf{c}, \mathbf{e}^{S})
  \label{eq_attn_sed}
\end{equation}

\subsection{Temp-tag-AAC}
\label{sec_tag_guided}


In our proposed temp-tag-AAC system, we transform the SED outputs into quantized temporal tags to make it easier for the model to learn the correspondence between SED outputs and captions.

We use double threshold post-processing~\cite{dinkel2021towards} with a low threshold of 0.25 and a high threshold of 0.75 to obtain the on- and off-sets of detected sound events from probability $\tilde{\mathbf{e}}^{S}$.
To infer relation between two different audio events, we compare the \textit{overlap} of them and the \textit{duration} of the shorter event. 
If \textit{overlap} is less than half of \textit{duration}, these two events are considered to occur sequentially; otherwise, they are considered to occur simultaneously. 
Based on the relations in audio clip obtained above, a 4-scale temporal tag representing the complexity of temporal information is extracted according to \Cref{tab_tamporal_tag}.
The process is shown as in \Cref{alg_process_sed}.
During \textbf{inference}, the temporal tag \textbf{inferred from the SED outputs} is used as $w_0$ fed to the decoder as temporal guidance, replacing the original $<$BOS$>$.

To help the model learn the correspondence between the temporal tag and the temporal descriptions in captions better, the \textbf{ground truth} tag is fed to the decoder during \textbf{training}.
The ground truth tag is extracted from the annotations based on the occurrence of conjunction words according to \Cref{tab_tamporal_tag}. 
We manually collect these conjunction words by analyzing the existing AAC datasets, such as ``while'', ``and'', \textit{etc.} indicating ``simultaneously,'' and ``follow", ``then'', \textit{etc.} indicating ``sequentially''.

\begin{algorithm}[htpb]
  \SetAlgoLined
  
  \KwData{Predicted probability $\tilde{\mathbf{e}}^{S}$}
  \KwResult{Temporal tag}
    $Relations \gets$ empty list\;
    $\{E: E_{on}, E_{off}\} \gets \mathrm{DOUBLE\_THRES}(\tilde{\mathbf{e}}^{S}) $\;
  \For{every pair A, B in \{E\} where $A_{on}<B_{on}$}{
    \textit{Overlap} $\gets A_{off}\mathrm{-}B_{on}$\;
    \textit{Duration} $\gets \mathrm{MIN}(A_{off}\mathrm{-}A_{on},B_{off}\mathrm{-}B_{on})$\;
    \eIf{$Overlap < 0.5 \times Duration$}{
        $Relations$.insert($``sequential"$)\;        
    }{
        $Relations$.insert($``simultaneous"$)\;
    }
  }
$Tag \gets$ Query \Cref{tab_tamporal_tag} using $Relations$\;   
Return $Tag$\;
  \caption{Infer temporal tags from SED results.} 
  \label{alg_process_sed}
\end{algorithm}

\begin{table*}[t]

  \caption{Results of system performance. ``$\text{FENSE}_{p=0}$'' indicates not penalizing grammatical errors in FENSE.}
  \label{tab_result1}
  \centering
  \begin{tabular}{ c| c|| c c c c c c| c c}
   \toprule
    Dataset & System & $\text{BLEU}_4$ & $\text{ROUGE}_\text{L}$ & CIDEr & METEOR & 
    SPICE & $\text{FENSE}_{p=0}$ &
    $\text{ACC}_{temp}$ & $\text{F1}_{temp}$\\
    \midrule
    
    \multirow{4}{*}{AudioCaps} & Baseline 
    &26.3 &\textbf{49.4} &\textbf{72.6} 
    &23.9 &17.6 &62.3 
    &37.3 &23.4\\

    &Cat-prob-AAC  &25.8 &\textbf{49.4} &71.2 
    &24.1 &17.5 &\textbf{62.6 }
    &39.2 &27.1\\
    
    &Attn-prob-AAC  &26.3 &49.1 &72.1 
    &23.9 &17.2 &62.3 
    &40.7 &30.4\\

    &Temp-tag-AAC (Ours)
    & \textbf{28.8} &\textbf{49.4} &68.2 &\textbf{25.5} &\textbf{18.5} &61.6 &\textbf{66.5} &\textbf{75.2}\\
    \midrule
    
    \multirow{4}{*}{Clotho} & Baseline 
    & \textbf{16.8} &\textbf{38.3} &\textbf{40.8} &\textbf{17.5} &\textbf{12.1} &48.4 &69.2 &17.4\\

    &Cat-prob-AAC 
    & 16.2 &38.0 &38.8 &17.3 &11.9 &\textbf{48.6} &\textbf{69.7} &17.7\\

    &Attn-prob-AAC
    & 16.7 &37.7 &38.7 &17.1 &11.9 &48.2 &\textbf{69.7} &16.4\\

    &Temp-tag-AAC (Ours)
    & 13.3 &35.3 &33.8 &15.9 &10.9 &47.2 &59.0 &\textbf{46.5}\\
    \bottomrule
  \end{tabular}
\end{table*}

\section{Experimental Setup}

\subsection{Datasets}
\label{sec_dataset}

AudioSet~\cite{gemmeke2017audioset} is a large-scale weakly-annotated sound event dataset, where sound events appearing in each audio clip are annotated, which consists of 527 categories.
AudioSet also provides a small-scale strongly-annotated subset~\cite{hershey2021benefit} which contains additional on- and off-sets of present events. 

AudioCaps~\cite{kim2019audiocaps} is the current largest AAC dataset, containing 50k+ audio clips collected from AudioSet. 
According to the extraction method mentioned in \Cref{sec_tag_guided}, 13487, 29399, 5438 and 8472 captions in AudioCaps annotations belong to the 4 scales respectively. 
The latter two more complex scenarios account for $\approx \frac{1}{4}$ of the total.

Clotho~\cite{drossos2020clotho} is another AAC dataset, containing 5k+ audio clips. 
The distribution of ground truth tag numbers in Clotho annotations is
8246, 18077, 926, 2396 respectively.
The latter two scales account for $\approx \frac{1}{10}$, which is much more imbalanced than AudioCaps.
As a matter of fact, Clotho derives from Freesound, where the audio clips often contain only the indicated sound with minimal background noise \cite[p.~51]{turpault:tel-03304880}\cite{martin2021diversity}.

\subsection{Hyper-parameters}

The SED model is first pre-trained on the weakly-annotated AudioSet, and then fine-tuned on the strongly-annotated AudioSet subset~\cite{hershey2021benefit}.
It achieves a $d$' of 2.37 on the strongly-annotated AudioSet evaluation set, compared with 1.39 in \cite{hershey2021benefit}, indicating that it provides reliable results for caption generation.

The training of audio captioning models, including the baseline model and other three approaches, follows the setup in \cite{xu2022sjtu}.
Models are trained for 25 epochs. 
Cross-entropy loss is used along with label smoothing ($\alpha = 0.1$).
We use a linear warm-up and an exponential decay strategy to schedule the learning rate, whose maximum value is $5\times 10^{-4}$.
Scheduled sampling is used with the proportion of teacher forcing decreasing linearly from $1$ to $0.7$.
Beam search with a size of 3 is adopted during inference.

\subsection{Metrics}

Generated captions are evaluated by our proposed temporal metrics and commonly-adopted metrics in AAC task.

\textbf{Temporal Metrics}
To better evaluate whether a generated caption includes temporal relations, we take the time-related conjunction words as a clue.
Captions can be classified upon whether there exists sequential conjunction words or not. 
The conjunction words include ``follow, followed, then, after'', which are only used to suggest temporal relations between sound events.
For example, ``Door closed \textit{then} a man talking" is regarded as a positive example (with temporal output) since ``then'' appears.
We exclude simultaneous conjunction words (e.g. ``and, with, as, while'') because they might carry semantic conjunction function and does not always signify temporal relations.
Whether these words express temporal relations cannot be recognized automatically and accurately.
Naturally, the temporal evaluation can be regarded as a binary classification evaluation: to determine whether there are sequential conjunction words or not in a caption.
We therefore use the binary classification evaluation metrics $\text{ACC}_{temp}$ and $\text{F1}_{temp}$ to measure the accuracy of temporal relation description. 
Within 5 reference captions, the maximum value contains the most detailed information and is taken as label to ensure the metric correctness.


\textbf{Overall Quality Evaluation Metrics}
We also adopt common audio captioning metrics to evaluate the overall quality of generated captions, including BLEU~\cite{kishore2002bleu}, $\text{ROUGE}_\text{L}$~\cite{chin-yew2004rouge}, METEOR~\cite{lavie2007meteor}, CIDEr~\cite{vedantam2015cider}, SPICE~\cite{anderson2016spice} and FENSE~\cite{zhou2022can}.
For FENSE we do not penalize grammatical errors to focus on evaluating the accuracy of captions' semantic information.

\section{Results and Analysis}


\subsection{Temporal Relation Enhancement}
Comparing temp-tag-AAC with the baseline model, our tag mechanism greatly improves the accuracy of temporal expressions on both datasets (shown in \Cref{tab_result1}).
For AudioCaps, both $\text{ACC}_{temp}$ and $\text{F1}_{temp}$ are significantly improved, suggesting the effectiveness of our method in enhancing temporal relations.
Due to the imbalanced categories of Clotho, $\text{F1}_{temp}$ is more reliable compared with $\text{ACC}_{temp}$, which also indicates a better capability to generate temporal-rich captions.

\begin{figure}[th]
  \centering
  \includegraphics[width=\linewidth]{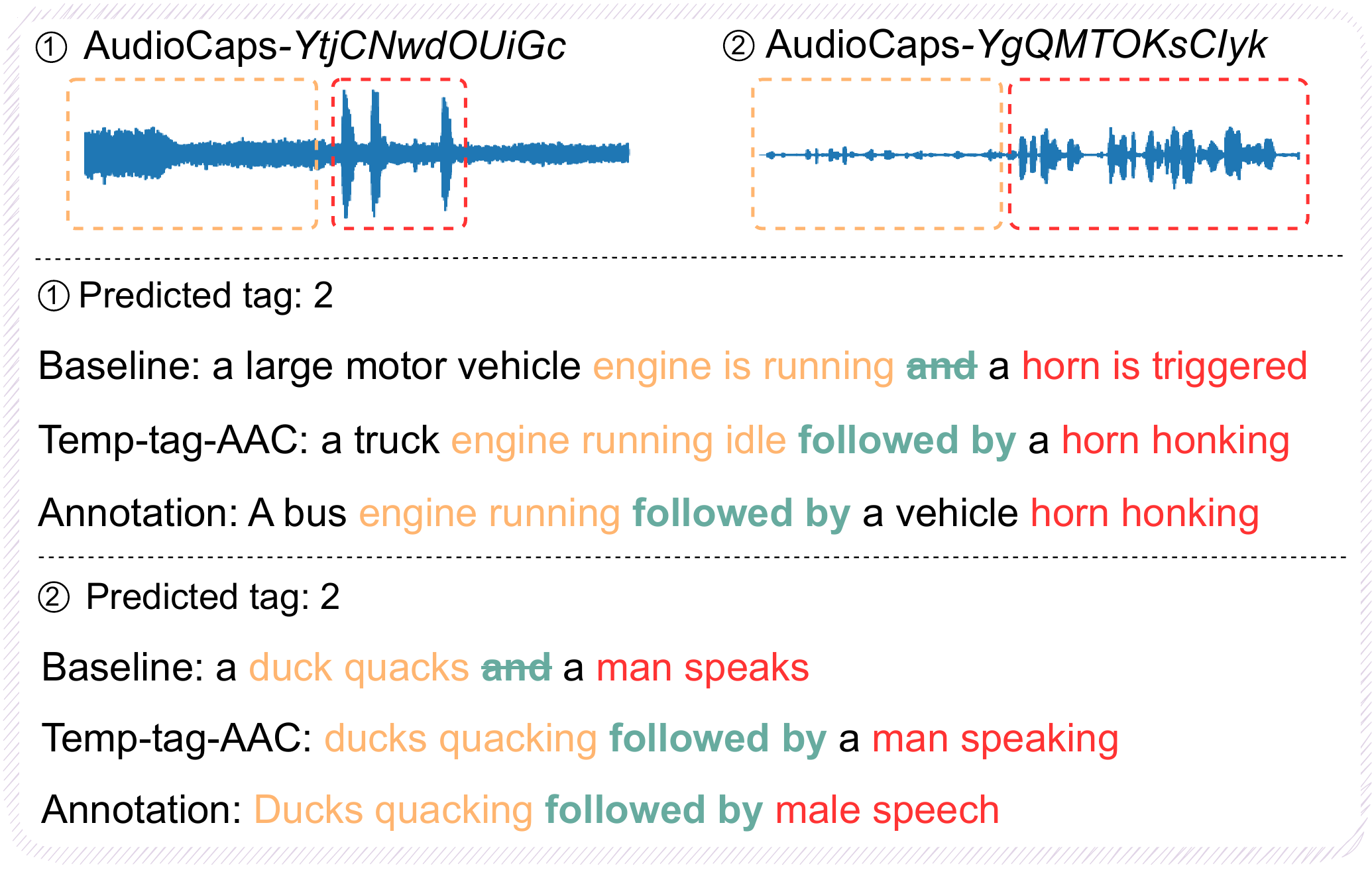}
  \caption{Output examples generated by baseline systems and tag guided approach. }
  \label{fig_sample}
\end{figure}

Without guidance, the baseline model tends to output general conjunction words that do not represent specific relations (``and'' and ``while'' are typical examples), resulting in loss of attention to the temporal relations between sound events.
Temp-tag-AAC restricts the output by inputting a tag which guides the model to use conjunction words to describe temporal relations. 
Typical examples are shown in \Cref{fig_sample}:
by incorporating the temporal tags, temp-tag-AAC successfully expresses the temporal relations while the baseline model simply uses ``and''. 

\subsection{Overall Quality of Generated Sentences}

The overall quality is evaluated by commonly-adopted metrics and shown in columns 3 to 8 of \Cref{tab_result1}.
On AudioCaps, temp-tag-AAC outperforms the baseline model on some metrics, but falls behind on others, indicating that our method is comparable to the baseline.
However, on Clotho, the quality of the caption sentences decreases, though the accuracy of temporal relations still sees an increase.
The performance drop is attributed to the data discrepancies between AudioSet and Clotho.
As stated in \Cref{sec_dataset}, Clotho audio samples exhibit vastly different characteristics from those in AudioSet.
As a result, the SED model trained on Audioset tends to output complex temporal tags (i.e., ``3'') for Clotho data when only one sound event is present.
The captioning model trained with such tags is prompted to generate sentences with complex conjunctions for single-event audios, which undermines its ability in generating reference-alike captions. 
The declined quality on Clotho indicates that adaptive SED deserves further exploration for generalization purpose.


\subsection{Comparison Between Different Approaches}

Comparing three different methods of integrating temporal information, we can conclude that direct integration by concatenation or attention only slightly improves the temporal description accuracy, but are far less effective than temp-tag-AAC.
This validates our intuition that human-like quantized prompts are more conducive to learning the correspondence between temporal information and conjunctions than direct outputs of SED.

\section{Conclusions}
This paper aims to improve the performance of expressing temporal information in AAC task.
We demonstrate that direct integration of SED outputs provides little help in improving the temporal relation description accuracy.
To overcome such challenge, we propose temp-tag-AAC which mimics human judgment by introducing 4-scale tags to guide the model to utilize temporal information.
Binary classification metrics $\text{ACC}_{temp}$ and $\text{F1}_{temp}$ are proposed to measure the accuracy of the temporal relation description.
Experimental results show that temp-tag-AAC significantly improves the temporal relation description accuracy.
With the guidance from the temporal tag, temp-tag-AAC uses conjunctions to express the temporal relations between sound events.
It is also comparable with the baseline in terms of the overall semantic quality of generated captions.



\bibliographystyle{IEEEtran}
\bibliography{mybib}

\end{document}